\documentclass[]{aa}
\usepackage{graphicx}
\usepackage{txfonts}

\newcommand{\xmm}{{\it XMM-Newton}}
\newcommand{\cha}{{\it Chandra}}
\newcommand{\rxte}{{\it RXTE}}
\newcommand{\bep}{{\it BeppoSAX}}
\newcommand{\asc}{{\it ASCA}}
\newcommand{\inte}{{\it INTEGRAL}}

\newcommand{\Halp}{H${\alpha}$}

\begin{document}

\title{Exploring the connection between the stellar wind and the non-thermal emission in LS~5039\thanks{The \Halp\ results presented here are based on observations made with
ESO Telescopes at the La Silla or Paranal Observatories under program IDs 67.D-0229(A), 69.D-0628(A) and 075.D-0591(A); the Observatoire de Haute-Provence; the
Observat\'orio do Pico dos Dias / LNA, Brazil; the G.D. Cassini telescope operated at the Loiano Observatory by the Osservatorio Astron\'omico di Bologna; and the Nordic
Optical Telescope, operated on the island of La Palma jointly by Denmark, Finland, Iceland, Norway, and Sweden, in the Spanish Observatorio del Roque de los Muchachos of the
Instituto de Astrofisica de Canarias (those data were taken using ALFOSC, which is owned by the Instituto de Astrof\'{\i}sica de Andaluc\'{\i}a (IAA) and operated at the
Nordic Optical Telescope under agreement between IAA and the NBIfAFG of the Astronomical Observatory of Copenhagen).}}

\author{V. Bosch-Ramon\inst{1} \and  C. Motch\inst{2} \and M. Rib\'o\inst{3}
\and R. Lopes de Oliveira\inst{2,4} \and \\  E. Janot-Pacheco\inst{4} \and 
I. Negueruela\inst{5} \and J.~M. Paredes\inst{3} \and A. Martocchia\inst{6}} 

\institute{Max Planck Institut f\"ur Kernphysik, Saupfercheckweg 1, Heidelberg 69117, Germany; 
vbosch@mpi-hd.mpg.de
\and Observatoire Astronomique, rue de l'Universit\'e 11, Strasbourg 67000, France;
motch@astro.u-strasbg.fr
\and 
Departament d'Astronomia i Meteorologia, Universitat de Barcelona, c/Mart\'i i Franqu\`es 1, 
08028 Barcelona, Catalonia, Spain; mribo@am.ub.es, jmparedes@ub.edu
\and 
Universidade de S\~ao Paulo, Instituto de Astronomia, Geof\'isica e Ci\^encias Atmosf\'ericas - IAG, Departamento de
Astronomia, Rua do Mat\~ao, 1226 - 05508-900 S\~ao Paulo, Brazil; rlopes@astro.iag.usp.br, janot@astro.iag.usp.br
\and 
Dpto. de F\'isica, Ingenier\'ia de Sistemas y Teor\'ia de la Se\~nal, Universidad de Alicante, Apdo. 99, 03080 Alicante, Spain: 
ignacio@dfists.ua.es
\and 
Istituto Nazionale di Astrofisica, Via del Fosso del Cavaliere 100, 00133 Roma, Italy; andrea.martocchia@iasf-roma.inaf.it}

\offprints{ \\ \email{vbosch@mpi-hd.mpg.de}}

\abstract
{LS~5039 has been observed 
with several X-ray instruments so far showing quite steady emission in the long term and no signatures of accretion disk. The source also presents X-ray 
variability at orbital timescales in flux and
photon index. The system harbors an O-type main sequence star with moderate mass loss. At present, the 
link between the X-rays and the stellar wind is unclear.}
{We aim here 
at studying the X-ray fluxes, spectra, and absorption properties of LS~5039 at apastron and periastron passages during an epoch of enhanced 
stellar mass loss, and the long term evolution of the latter in connection with the X-ray fluxes.}
{New \xmm\ observations have been performed around periastron and apastron passages in September 2005, 
when the stellar wind activity was apparently higher. April 2005 \cha\ observations on LS~5039 are also revisited. 
Moreover, a compilation of \Halp\ EW data obtained since 1992, from which the stellar 
mass loss evolution can be approximately inferred, is carried out.}
{\xmm\ observations show higher and harder emission around apastron than around periastron. No signatures of thermal emission 
or a reflection iron line indicating the presence of an accretion disk are found in the spectrum, 
and the hydrogen column density ($N_{\rm H}$) is compatible with being the same in both observations and consistent 
with the interstellar value. 
2005 \cha\ observations show a hard X-ray spectrum, and possibly high fluxes, although pileup effects preclude obtaining conclusive results. 
The \Halp\ EW shows
yearly variations of a $\sim 10$\%, and does not seem to be correlated with X-ray fluxes obtained at similar phases, 
unlike it would be expected in the wind accretion scenario.}
{2005 \xmm\ and \cha\ observations are consistent with 2003 \rxte/PCA results, namely moderate flux and spectral variability at different orbital 
phases. The constancy of the $N_{\rm H}$ seems to imply that either the X-ray emitter is located at $\ga 10^{12}$~cm from the compact object, 
or the real density in the system is 3 to 27 times smaller than the one predicted by a spherical symmetric wind model. 
We suggest that the multiwavelength non-thermal emission of LS~5039 is related to the observed extended radio jets and
unlikely produced inside the binary system.}

\maketitle

\section{Introduction} \label{intro}

LS~5039 (V479 Sct) is an X-ray binary (Motch et~al. \cite{motch97}) located at 2.5$\pm 0.5$~kpc (Casares et~al. \cite{casares05}; C05 hereafter). The orbital
ephemeris of the source have been obtained by C05, being: $T_0\,=\,$HJD~$2451943.09\pm 0.1$, periastron passage -phase 0.0-; $P_{\rm orb}=3.90603\pm 0.00017$~days, orbital
period; $e=0.35\pm 0.04$, eccentricity; and $i\sim 15^{\circ}$--$60^{\circ}$, inclination\footnote{We note that all through this work, the errors in phases -not
shown- range between 0.03--0.04 due to  the uncertainties in $T_0$ and $P_{\rm orb}$.}. LS~5039 presents -$\sim$1--100~milliarcseconds- radio jets (Paredes et~al.
\cite{paredes00,paredes02}), shows X-ray variability on timescales similar to the orbital one (Bosch-Ramon et~al. \cite{bosch05}; BR05 hereafter), and has been
detected at very high-energy gamma-rays (Aharonian et al. \cite{aharonian05}), which virtually confirms its association with a {\it CGRO}/EGRET (EGRET from now on)
source (Paredes et~al. \cite{paredes00}). Interestingly, the TeV emission varies with the orbital period (Aharonian et al. \cite{aharonian06}). Recently, LS~5039
has been also detected by \inte\ (Goldoni et~al. \cite{goldoni06}) up to 100~keV (De Rosa et~al. \cite{derosa06}).  
The nature of the compact object is still uncertain. C05
suggest that it may be a black hole, although there is an on-going debate on this issue, and some authors have proposed that LS~5039 is in fact a young
non-accreting pulsar (see, e.g., Martocchia et~al. \cite{martocchia05} -M05 hereafter-; Dubus \cite{dubus06}).

All X-ray observations performed so far by imaging instruments failed to show strong flux and photon index variations. \rxte/PCA (\rxte\ from now on) observed
also the source in 1998 (Rib\'o et~al. \cite{ribo99}) and 2003 (BR05), although the latter authors found that \rxte\ data were likely contaminated by diffuse X-rays
from the Galactic Ridge. Still, 2003 \rxte\ data showed flux variations by a factor of $\sim 2$ unrelated with diffuse background emission, peaking smoothly around
phase 0.8 and more sharply at other phases. These peaks were apparently accompanied by spectral hardening.  At phases $\sim 0.8$, higher and harder emission has
been also observed at TeV energies (Aharonian et~al. \cite{aharonian06}), and simultaneous \cha\ observations have apparently shown a similar behavior
in X-rays (Horns et~al. \cite{horns06}). A possible relationship between the mass-loss rate of the companion star and the X-ray emission in LS~5039 was discussed
in previous works and a correlation suggested in the context of wind accretion powering the X-rays (e.g., Reig et~al. \cite{reig03} -R03 hereafter-; McSwain et~al.
\cite{mcswain04} -MS04 hereafter-). 

In this work, we present recent \xmm\ observations of LS~5039 carried out in September 2005 during periastron and apastron passages, in an epoch of inferred high
stellar mass-loss rate. \cha\ data taken in the same year is reanalyzed. In addition, we investigate the long-term \Halp\ EW evolution, linked to the
evolution of the stellar mass-loss rate, and compare it with the X-ray fluxes looking for possible correlations. These results are put in context with information
at other wavelengths and a physical scenario is suggested.

\section{X-ray observations}\label{xobs}

\subsection{\xmm\ observations}

The \xmm\ observations presented here were performed in two runs, on the 22th and 24th of September 2005, with EPIC pn on times of 15.8~ks and 10.4~ks (very similar
to those  of MOS1 and MOS2), respectively (see Table~\ref{tab0}). The corresponding observed phases are 0.49--0.53, around apastron, and 0.02--0.05, right after
periastron. These observations were triggered in the context of a target of opportunity (ToO) program linked to the strength of the \Halp\ line, which gives
information on the state of the stellar wind. The \Halp\ EW was found in late August 2005 to be above of $-$2.4\AA, defined as the threshold set to trigger the
\xmm\ observations and significantly larger than the values observed during the previous  years. \xmm\ observations were carried out in
September due to visibility constraints. The physical motivation of the ToO was that, if direct accretion of the stellar wind is taking place in the system, the
detection of higher X-ray fluxes, and/or accretion spectral features like a multi-color black-body and/or a reflection iron line, seems more likely during epochs
of larger stellar mass loss, i.e. higher accretion rate.  The convention for the \Halp\ EW in this work is to take this value as negative when in absorption, i.e.
it increases in the same sense as the stellar mass-loss rate.

 \begin{table}[]
  \caption[]{X-ray observations and spectral results.}
  \label{tab0}
  \begin{center}
  \begin{tabular}{llllll}
  \hline\noalign{\smallskip}
  \hline\noalign{\smallskip}
Date & 2005-04-13 & 2005-09-22 & 2005-09-24\\
Mission & \cha & \xmm & \xmm \\
OBSID & 6259 & 0202950201 & 0202950301 \\
Orbital phase & 0.86--0.87 & 0.49--0.53 & 0.02--0.05 \\
MJD$_{\rm start}$ & 53473.116 & 53635.714 & 53637.779 \\
MJD$_{\rm stop}$ & 53473.174 & 53635.897 & 53637.899 \\
$f_{x}$ ($L_{x}$) & 2$\pm 1$ (1.5) & 1.18$^{+0.03}_{-0.07}$ (0.9) &
0.74$\pm 0.07$ (0.6)\\
Photon index & 1.1$\pm 0.2$ & 1.51$\pm 0.02$ & 1.59$\pm 0.03$ \\
$N_{\rm H}$ ($phabs$) & 6$\pm 1$ & 6.3$\pm 0.1$ & 6.2$\pm 0.2$ \\
  \noalign{\smallskip}\hline
  \end{tabular}
  \end{center}
\begin{list}{}{}
\item Notes: fluxes and luminosities are given unabsorbed in the
0.3--10 keV energy band in units of $10^{-11}$~erg~cm$^{-2}$~s$^{-1}$
and $10^{34}$~erg~s$^{-1}$, respectively.  The $N_{\rm H}$ units are 
$10^{21}$~cm$^{-2}$.
Quoted errors are at the
1~$\sigma$ confidence level. Ephemeris taken from C05.
\end{list}
\end{table}

We focus here on the data obtained with the EPIC instrument, since the X-ray spectrometer (RGS) data did not have enough statistics due to the faintness of the
source. The MOS and pn cameras were in small and full window respectively, with medium filters. The reduction and analysis of the data have been performed
following the standard procedure using the SAS 7.0.0 software package to filter data and create spectra and lightcurves, and Xspec 12.3.0 for the spectral
analysis. No large background flare was present in any of the two data sets.  The 0.3--12~keV count rates per camera around phases 0.5 and 0.0 were 0.46 and 0.29~cts~s$^{-1}$ for MOS1, 0.47 and 0.30~cts~s$^{-1}$ for MOS2,
and 1.09 and 0.71~cts~s$^{-1}$ for pn, respectively. 

\begin{figure}[t]
\resizebox{\hsize}{!}{\includegraphics{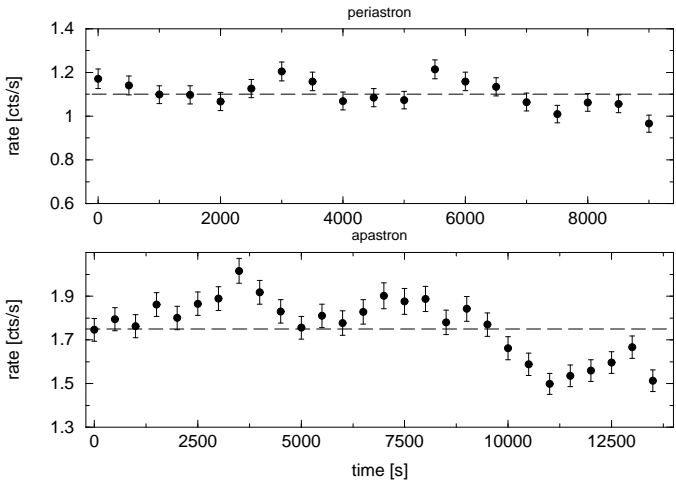}}
\caption{Lightcurves at periastron (top) and apastron (bottom). The average value is given as a long-dashed line. We note that the 
time axis range differently in each lightcurve.}
\label{lc0/0.5}
\end{figure}

The lightcurves for the two runs, adding the data from pn, MOS1 and MOS2 cameras and taking time bins of 500~s, are presented in Fig.~\ref{lc0/0.5}  for
periastron (bottom) and apastron (top) passages, respectively. Around apastron, the emission is variable on timescales of a few hours. A constant fit to the data
gives a reduced $\chi$ squared ($\chi_{\rm red}^2$; 27 degrees of freedom -d.o.f.-) of 6.8, and the maximum count rate variation is $\approx 25$\% during this run,
with deviations of $\approx \pm 5\sigma$ from the mean value. 
Around periastron, the lightcurve is still somewhat
inconsistent with being constant ($\chi_{\rm red}^2=2.4$; 18 d.o.f.), with a maximum count rate variation of $\approx 25$\% and deviations of 
$\approx \pm 2\sigma$ from the mean value. We
studied also the hardness ratio evolution for both runs, which is compatible with being constant. Searching for pulsations, no significant period peaks are found  in the frequency range of 0.01 to 83 Hz
and the pulsed fraction has been constrained to be less than $\approx 10$\%, similar to the results reported by M05 for 2003 \xmm\ data.

\begin{figure}[]
\includegraphics[angle=0, width=0.45\textwidth]{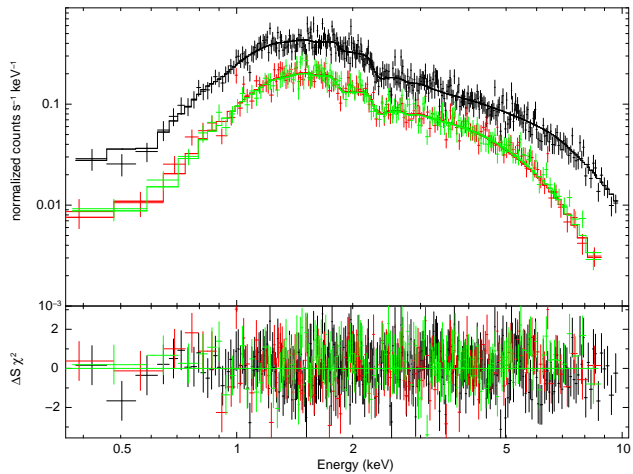}
\caption{Spectral data from pn (upper spectrum), MOS1 and MOS2 (lower spectra) cameras, and the best 
fit to the three of them: an absorbed power-law, around phase 0.5 (those around phase 0.0 look very similar and are not shown here). 
The deviations in $\sigma$ of the model from the data are shown in the lower panel.}
\label{spec05}
\end{figure}

We show in Fig.~\ref{spec05} the spectrum of the source around phase 0.5, altogether with the deviations in $\sigma$ of the best fit model from the data. We
have added the data obtained from the pn, MOS1 and MOS2 cameras. The spectrum around phase 0.0 is not shown though it looks very similar to the one around
phase 0.5. The spectrum of the emission is well represented by an absorbed power-law ($\chi_{\rm red}^2=1.08$ -637 d.o.f.- and 0.93 -312 d.o.f.- for phases 0.5 and 0.0,
respectively). The unabsorbed fluxes, photon indices (0.3--10~keV) and $N_{\rm H}$ values for the two runs are given in Table~\ref{tab0}. The flux between
periastron and apastron increases by a 60\% at a level of 4.4$\sigma$, and the photon index rises up marginally by 0.08 (2.2$\sigma$).
This behavior,
despite of being marginal, give support to the \rxte\ finding that the higher the flux, the harder the spectrum. For the absorption model, we used $phabs$, which assumes updated photoelectric cross-sections from Balucinska-Church \& Mc~Cammon
(\cite{balu92}). The $N_{\rm H}$ values are compatible within errors between both runs. 

A multi-color black-body ($bbody$) and a Gaussian line component have been added, separately, to the absorbed power-law model for the two runs. We obtain
upper-limits for the multi-color black-body flux in the two runs similar to those found by M05. At the same phases, we
look for flux upper-limits of a possible iron line 
and obtain an
upper-limit for the flux of about a 2\% the total one. The lack of a multi-color black-body and a line component in the data is consistent with previous
imaging instrument observations and confirms that the iron line reported in Rib\'o et~al. (\cite{ribo99}) was due to background contamination.

\subsection{\cha\ data}

In April 2005, \cha\ observed LS~5039 in the ACIS-S mode (time frame: 1.6~s) during 5~ks (see Table~\ref{tab0}) at a time  
corresponding to the orbital phase 0.86--0.87, at which \rxte\ observed a smooth maximum (BR05). The \cha\ Interactive Analysis of Observations software
package (CIAO~3.3.0.1) has been used to extract the spectrum and the lightcurve, following the standard procedure given in the \cha\ analysis threads.  {\it Sherpa}
(CIAO~3.3) has been used to analyze the spectrum. The ACIS count rate was relatively high, 0.3~cts~s$^{-1}$ (0.3--10~keV). \cha\ data is likely moderately
affected by pileup (see {\it The Chandra Proposers' Observatory Guide}, Fig.~6.19). Therefore, we have used the pileup model available in Sherpa to obtain our fits
(Davis \cite{davis01}).

The spectrum can be well fitted by a power-law model with absorption ({\it xsphabs}, the sherpa version of {\it
phabs}) and pileup. We perform the spectral fit fixing the event pileup fraction parameter to 0.95 to avoid unreasonably high values for the flux (i.e.
$\sim 10^{-10}$~erg~cm$^{-2}$~s$^{-1}$). We obtain a relatively low grade migration parameter of $0.2\pm 0.1$. The pileup fraction is $\approx 11$\%.
Therefore, the effect of pileup on these observations seems to be only moderate. 
The results of our fit are given in Table~\ref{tab0}. The photon index value, harder than that found in 
almost all previous observations, has been also calculated without a pileup model 
by removing the central pixels, i.e. the most affected by pileup (pileup effects can indeed produce an artificial hardening of the spectrum), 
obtaining compatible spectral results. No disk or line components are required to fit the data. The lightcurve is consistent with being constant.

\section{\Halp\ line observations} 

\begin{table*}[]
\begin{center}
\caption{\Halp\ EW measurements obtained from 1992 to 2006$^*$.}
\begin{tabular}{lccllrlcc}
\hline\noalign{\smallskip}
\hline\noalign{\smallskip}
Date & MJD & Orb. phase & Observatory & Instrument & Resolution & Setting$^{**}$ &
\Halp\ EW & Error \\
(yyyy-MM-dd)  &  &  &  &  & & &(\AA) & (\AA) \\
\hline\noalign{\smallskip}
1992-04-19     &     48731.250    &    0.85    &    ESO-2.20    &    EFOSC2
    &     3,000         &  -   &      $-$2.80        &        0.07 \\
1993-06-19     &     49257.010    &    0.45    &    OHP-193     &
CARELEC    &     450$^{*}$      &  -   &      $-$2.50        &
0.10\\
1999-08-18     &     51408.842    &    0.35    &    OAB        &     BFOSC
    &    1,600          & Grism\#7    &     $-$2.60        &
0.10\\
2000-09-08     &     51795.813    &    0.42    &    OHP-193     &
 ELODIE    &     45,000      & -    &       $-$3.24        &
0.10\\
2000-09-10     &     51797.768    &    0.92    &    OHP-193     &
 ELODIE    &     45,000      & -    &       $-$2.91        &
0.15\\
2000-09-12     &     51799.768    &    0.43    &    OHP-193     &
 ELODIE    &     45,000      & -    &       $-$3.23        &
0.15\\
2000-10-18     &     51835.984    &    0.70    &    ESO-1.52     &    FEROS
    &     48,000     & -    &       $-$2.64        &          0.05\\
2000-10-19     &     51836.983    &    0.96    &    ESO-1.52     &    FEROS
    &     48,000     & -    &      $-$2.61        &          0.05\\
2000-10-20     &     51838.000    &    0.22    &    ESO-1.52     &    FEROS
    &     48,000     & -    &      $-$2.71        &          0.05\\
2001-05-24     &     52053.424    &    0.37    &    ESO-VLT-U1    &    FORS1
    &      990         & GRIS\_600V    &      $-$2.69        &
0.07\\
2001-05-26     &     52055.417    &    0.88    &    ESO-VLT-U1    &    FORS1
    &      990         & GRIS\_600V    &      $-$2.69        &
0.07\\
2001-08-25     &     52146.858    &    0.29    &    OHP-193     &     ELODIE
&     45,000     & -    &       $-$2.95        &          0.10\\
2001-08-26     &     52147.857    &    0.55    &    OHP-193     &     ELODIE
&     45,000     & -    &       $-$2.98        &          0.10\\
2001-08-28     &     52149.848    &    0.06    &    OHP-193     &     ELODIE
&     45,000     & -    &       $-$2.91        &          0.20\\
2001-08-30     &     52151.822    &    0.56    &    OHP-193     &     ELODIE
&     45,000     & -    &       $-$2.92        &          0.20\\
2001-10-06     &     52188.978    &    0.08    &    ESO-1.52    &    FEROS
    &     48,000     & -    &      $-$2.64        &          0.10\\
2001-10-07     &     52189.992    &    0.33    &    ESO-1.52    &    FEROS
    &     48,000     & -   &      $-$2.79        &          0.10\\
2001-10-08     &     52191.050    &    0.60    &    ESO-1.52    &    FEROS
    &     48,000     & -    &      $-$2.50        &          0.15\\
2001-10-09     &     52192.000    &    0.85    &    ESO-1.52    &    FEROS
    &     48,000     & -    &      $-$2.53        &          0.10\\
2002-06-14     &     52439.340    &    0.17    &    ESO-NTT        &    EMMI
    &     5,000         & -    &      $-$2.81        &          0.05\\
2002-07-08     &     52463.335    &    0.31    &    ESO-NTT        &    EMMI
    &     5,000         & -    &      $-$2.78        &          0.05\\
2002-07-09     &     52464.141    &    0.52    &    ESO-NTT        &    EMMI
    &     5,000         & -    &      $-$2.80        &          0.05\\
2002-07-10     &     52465.052    &    0.75    &    ESO-NTT        &    EMMI
    &     5,000         & -    &      $-$2.72        &          0.10\\
2002-07-11     &     52466.214    &    0.06    &    ESO-NTT        &    EMMI
    &     5,000         & -    &      $-$2.70        &          0.05\\
2004-07-13     &     53199.923    &    0.89    &    OHP-193     &    CARELEC
&     900$^{*}$     & Grating 133/375    &      $-$2.60        &
      0.10\\
2004-07-14     &     53200.937    &    0.15    &    OHP-193     &    CARELEC
&     900$^{*}$     & Grating 133/375    &      $-$2.54        &
      0.10\\
2004-07-16     &     53202.936    &    0.66    &    OHP-193     &    CARELEC
&     900$^{*}$     & Grating 133/375    &      $-$2.44        &
      0.10\\
2004-07-18     &     53204.906    &    0.17    &      OAB          &
BFOSC   &     2,000         & Grism\#8    &      $-$2.47        &
0.10 \\
2004-09-02     &     53250.935    &    0.95    &    LNA-0.60    &
ESPCASS    &     1,460         & -    &      $-$2.60        &
0.10\\
2005-04-07     &     53467.299    &    0.34    &    LNA-1.60    &    ESPCASS
&     3,600         & Grating 1200/750    &      $-$2.83        &
0.10\\
2005-04-07     &     53467.299    &    0.34    &    LNA-1.60    &    ESPCASS
&     3,600         & Grating 1200/750    &      $-$2.77        &
0.10\\
2005-07-31     &     53583.748    &    0.16    &    ESO-NTT        &
EMMI      &     1,100         & Grism\#5    &      $-$2.55        &
0.05\\
2005-08-20     &     53602.935    &    0.07    &    LNA-1.60    &    ESPCASS
&     2,500         & Grating 900/550    &       $-$2.39        &
0.05\\
2005-08-20     &     53602.963    &    0.08    &    LNA-1.60    &    ESPCASS
&     2,500         & Grating 900/550    &      $-$2.34        &
0.05\\
2005-08-21     &     53603.078    &    0.10    &    LNA-1.60    &    ESPCASS
&     2,500         & Grating 900/550    &      $-$2.43        &
0.05\\
2005-09-30     &     53643.005    &    0.33    &    LNA-1.60    &    ESPCASS
&     730         & Grating 300/640    &      $-$2.51        &
0.10\\
2006-05-07     &     53862.181    &    0.44    &      NOT        &
 ALFOSC    &     3,000         & Grism\#16    &      $-$2.40        &
0.10\\
\hline\noalign{\smallskip}
\label{halpobs}
\end{tabular}
\end{center}
{
$^*$ Spectra with EW values corrected for the presence of blended telluric
lines.\\
$^{**}$ The setting of the instrument is shown when relevant.
}
\end{table*}

%We list in Table~\ref{halpobs} the journal of optical observations and the corresponding \Halp\ EW. Most of these optical spectroscopic
%observations of LS 5039 were carried out by us at regular intervals of time between September 2000 and July 2002, with the aim to construct
%an orbital solution and to monitor the variations of the EW of the \Halp\ line, the most convenient line to study the wind of LS~5039. A new observing campaign was initiated in 2004 in order to determine the time at which the \Halp\ EW
%would go above $-$2.4~\AA, the condition required to trigger the 2005 \xmm\ observations. 

We present 
%as on-line material 
in Table~2
the list of the journal of optical observations and the corresponding \Halp\ EW. Most of these optical
spectroscopic observations of LS 5039 were carried out by us at regular intervals of time between September 2000 and July 2002, with the aim
to construct an orbital solution and to monitor the variations of the EW of the \Halp\ line, the most convenient line to study the wind of
LS~5039. A new observing campaign was initiated in 2004 in order to determine the time at which the \Halp\ EW would go above $-$2.4~\AA, the
condition required to trigger the 2005 \xmm\ observations. 

Various instruments were used in the course of this extended campaign. At the Observatoire de Haute-Provence (OHP), we observed most of the time with the ELODIE
Echelle Spectrometer. Some additional spectra were obtained with the CARELEC spectrograph
(Lemaitre et al. \cite{lemaitre90}). At ESO, the majority of the data were acquired with the FEROS\footnote{All FEROS observations were
obtained during the Brazilian time allocation.} Echelle Spectrometer (Kaufer et al. \cite{kaufer99}). FORS1 observed the source on two occasions in 2001. LS 5039 
was also monitored over an orbital cycle with the ESO-NTT and EMMI during a dedicated program in July 2002. Finally, the spectrum obtained in 1992 is described in Motch et al. (\cite{motch97}). The spectra obtained on 1999-08-18 and 2004-07-19 were acquired at Osservatorio Astronomico di Bologna with the 1.52~m Loiano telescope and the BFOSC instrument and at the NOT we used the ALFOSC spectrograph.
At the Brazilian National Observatory of the Observat\'orio do Pico dos Dias, we used the ESPCASS Cassegrain grating spectrometer mounted on either the 0.6~m or
the 1.6~m telescope. Instrument settings are summarized in the 
%Table presented on-line. 
Table 2. 

The \Halp\ line has an intrinsic FWHM of $\sim$8~\AA\ and is therefore relatively well resolved in most of our observations. Due to
the southern location of the object, observations from the OHP and Canary Islands were always acquired at a high air mass, typically
1.9 at OHP. At these high air masses, several telluric absorption lines start to be noticeable and can substantially change the value
of the \Halp\ EW, either by directly adding to the absorption profile or by changing the value of the
reference continuum. These lines are clearly seen in the high resolution ELODIE individual spectra. In the $\lambda\lambda$ range
6540--6580~\AA, the most important telluric lines are located at $\lambda\lambda$ 6543.89, 6546.61, 6552.62, 6557.16, 6564.19, 6572.06 and
6574.83~\AA. Guided by the atmospheric transmission lines atlas of Hinkle et~al. (\cite{hinkle03}), we could easily remove the sharp
lines due to telluric absorption from the high resolution ELODIE spectra and thus fully correct for atmospheric effects. Some of the
FEROS spectra also exhibited evidences of well resolved atmospheric lines at a much lower level than at OHP and could be removed as
well. 

At a resolution of $\sim$ 1000 or less, however, atmospheric lines are heavily smeared and cannot be simply removed. They appear as shoulders
in the blue and red wings of the broad \Halp\ absorption line of LS 5039 or as slight distortion in the central line profile. We
estimated the effect of these lines on the measured equivalent widths by scaling the absorption spectrum of Hinkle et~al.
(\cite{hinkle03}) so that they could match the equivalent widths of the narrow lines observed in the mean OHP ELODIE spectrum and
applying it to a Gaussian \Halp\ line profile fitting that of LS 5039. Because of the low altitude and relatively high humidity level, 
the atmospheric effects at OHP are noticeably large. We find that telluric absorptions can easily add 0.4~\AA\ to
the equivalent width of a low resolution spectrum at airmass 1.9. Low resolution spectra obtained at OHP were thus rectified
using continuum regions clear of known atmospheric lines and their EW were then corrected by 0.4~\AA. In the absence of clear measurements 
or estimates of water vapor column we did not try to correct the other measurements obtained at lower air masses and under much dryer conditions. 

\section{Discussion}\label{disc}

\subsection{Evolution of the \Halp\ EW and the X-rays}\label{xhr}

\begin{figure}[]
\resizebox{\hsize}{!}{\includegraphics{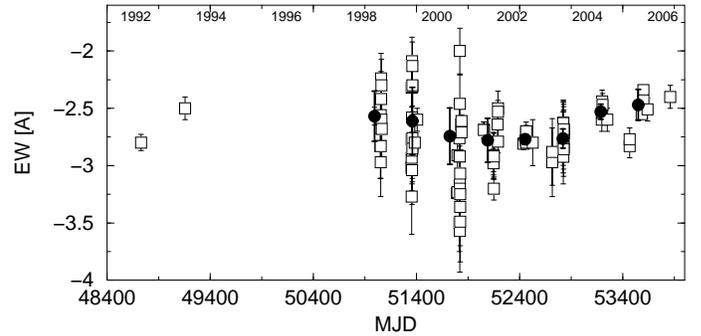}}
\caption{The year averages (black circles) and the individuals points (white squares) for the \Halp\ EW. For those years with several data points, the 
year average points are shown and their error bars represent the dispersion of values. The shown data come from 
the Table 2 
and previous works (C05: year 2003; MS04: years 1998, 1999, 2000, 2003; R03: years 1999, 2002).}
\label{alph}
\end{figure}

We have studied the available \Halp\ data (this work; C05; MS04; R03) and found, as already suggested in previous works (e.g. McSwain \& Gies
\cite{mcswain02}, R03), that  the \Halp\ EW is not compatible with being constant ($\chi_{\rm red}^2=3.7$; 96 d.o.f.), with a weighted mean
value of $-2.66\pm 0.01$~\AA, where the error has been computed using the error propagation method. All errors quoted here are 1~$\sigma$. The
weighted standard deviation of the sample is 0.21~\AA, giving an idea of the dispersion degree of the individual points. The yearly averages
with their corresponding standard deviations, altogether with the individual points, are shown in Fig.~\ref{alph}. For those years when only
one measurement is available, we show just the individual point with its error. The yearly mean errors are typically $0.02-0.1$ (not shown in
the plot), being significantly smaller than the yearly standard deviations  (shown in the plot), $0.04-0.29$~\AA. This relatively large
dispersion seems to indicate that the \Halp\ EW also varies at scales shorter than one year. Due to the heterogenous origin of the data, we
have done a further test to check whether the \Halp\ EW is indeed variable using only data given in the 
%Table presented on-line, 
Table 2, 
treated
homogeneously, excluding those points that may be affected by atmospheric lines, concretely those obtained with CARELEC at OHP, and the two
much earlier 1992 and 1993 points. The resulting mean is $-2.62\pm 0.01$~\AA\, with a standard deviation of 0.26, being also this data subset
incompatible with a constant ($\chi_{\rm red}^2=5.6$, 24 d.o.f.). Although the strong dispersion of the data prevents to make strong claims,
it is worthy noting that the \Halp\ EW seems to increase on average during the period 2003--2006, which would be evidence of a higher stellar
mass-loss rate.  

As noted in Sect.~\ref{intro}, a correlation between the \Halp\ EW and the X-ray flux has been proposed in the past (e.g. R03; MS04) suggesting that larger
stellar  mass loss rates are correlated with episodes of higher X-ray luminosity. These authors included in their comparison non imaging 1998 \rxte\ data,
and the used X-ray observations were taken at different orbital phases, being likely affected by additional orbital variability. Trying to avoid orbital
variation effects and excluding \rxte\ data, we use \xmm\ data taken in 2003 and 2005 around apastron and find that the flux in 2005 is only a $\approx 10$\%
larger ($2\sigma$ level) than that observed in 2003. Otherwise, the \Halp\ EW appears larger by a $\approx 14$\% ($6\sigma$ level; between $-2.76\pm
0.05$~\AA\ to $-2.43\pm 0.02$~\AA)\footnote{These has been obtained averaging few months data around the \xmm\ observation dates, which implies to assume a
steady \Halp\ behavior at these timescales.}, which could indicate variations of the stellar mass-loss rate by a factor of $\sim 2$ (see, e.g., Puls et al.
\cite{puls96}, Table 7). Therefore, similar phase/epoch X-ray/\Halp\ EW data point to a quite constant X-ray flux versus a significantly changing wind
state. This absence of clear X-ray flux response to a significant change in the stellar mass loss rate is 
at variance with the relation suggested by RO3 and MS04, although it is presently just a two-points comparison.

Finally, we show in Table~\ref{tab1} the 0.3--10~keV unabsorbed fluxes, photon indices and $N_{\rm H}$ (using the $wabs$ model) values obtained by all
instruments with imaging capabilities and working in the mentioned energy range. Using the values provided in that table, we obtain the following weighted
mean values for the 0.3--10~keV flux, $(1.05\pm 0.02)\times 10^{-11}$~erg~cm$^{-2}$~s$^{-1}$ (fit-to-constant $\chi_{\rm red}^2=8.4$; 7 d.o.f.), the photon
index, $1.54\pm 0.01$ ($\chi_{\rm red}^2=2.7$; 7 d.o.f.), and $N_{\rm H}$, $(6.8\pm 0.1)\times 10^{21}$~cm$^{-2}$ ($\chi_{\rm red}^2=0.7$; 7 d.o.f.).
Therefore, the flux found by all the instruments with imaging capabilities is clearly variable (dominated by orbital variability?) even when excluding the
2005 \cha\ data point, the photon index seem to change as well with time, and the $N_{\rm H}$ is clearly compatible with being constant.

 \begin{table*}[]
  \begin{center}
  \caption[]{Summary of the results for 
all the X-ray observations on LS~5039 performed by instruments with imaging capabilities. The
unabsorbed fluxes in the energy range 0.3--10~keV, photon indices, and $N_{\rm H}$ (with $wabs$ model) values are given.
\bep\ flux errors are the flux dispersion 
accounting for the count rate changes by a factor of two during the observation and neglecting 
photon index variations. \asc\ flux errors are assumed to be a 10\% of the flux. The given 2005 
\xmm\ $N_{\rm H}$ values are computed with the {\it wabs} model for comparison with 2003 values. 
For more details, see M05 (2003 \xmm\ data), 
R03 (\bep\ data), and Yamaoka, K., private communication (\asc\ data). 
\label{priorobs}}
  \label{tab1}
  \begin{tabular}{lllllll}
  \hline\noalign{\smallskip}
  \hline\noalign{\smallskip}
Date & MJD & Mission & Phase range & 
 Flux (0.3--10~keV) &Photon index & $N_{\rm H}$ ($wabs$) \\
yyyy-MM-dd & & & &$\times10^{-11}$~erg~cm$^{-2}$~s$^{-1}$ &  & $\times10^{21}$~cm$^{-2}$ \\
  \hline\noalign{\smallskip}
1999-10-04 & 51455.9 &\asc & 0.38--0.47 & 0.80$\pm 0.08$	    & 1.6$\pm 0.1$            & 7$\pm 1$ \\
2000-10-01 & 51818.5 &\bep & 0.97--0.21 & 0.5$\pm 0.2$	    & 1.8$\pm 0.2$            & 10$^{+3}_{-4}$ \\
2002-09-10 & 52527.3 &\cha  & 0.71--0.73& 0.8$\pm 0.3$	     & 1.1$\pm 0.2$           & 6$\pm 2$ \\
2003-03-08 & 52706.3 &\xmm  & 0.54--0.57& 1.03$\pm 0.07$	     & 1.56$^{+0.02}_{-0.05}$ & 7.2$^{+0.3}_{-0.5}$\\
2003-03-27 & 52725.8 &\xmm  & 0.55--0.58& 0.97$\pm 0.07$	     & 1.49$^{+0.05}_{-0.04}$ & 6.9$^{+0.5}_{-0.3}$\\
2005-04-13 & 53473.1 &\cha & 0.86--0.87 & 2$\pm 1$		    & 1.1$\pm 0.2$            & 6$\pm 1$ \\
2005-09-22 & 53635.8 &\xmm & 0.49--0.53 & 1.18$^{+0.03}_{-0.07}$& 1.51$\pm 0.02$           & 6.7$\pm 0.2$ \\
2005-09-24 & 53637.8 &\xmm & 0.02--0.05 & 0.74$\pm 0.07$	    & 1.59$\pm 0.03$          & 6.6$\pm 0.2$ \\
  \noalign{\smallskip}\hline
  \end{tabular}
  \end{center}
\end{table*}

\subsection{The $N_{\rm H}$ value and the location of the X-ray emitter}

Photo-electric absorption in the dense wind of the primary in LS~5039 should increase the observed $N_{\rm H}$ to values above that due to the interstellar medium
(ISM) alone. This effect could be detectable when comparing the values of the $N_{\rm H}$ inferred from X-ray and optical observations, respectively. However, the
$N_{\rm H}$ values inferred from \xmm\ data (using either $wabs$ or $phabs$; see Sect.~\ref{xobs}) are similar to, and somewhat smaller than, the value of the ISM
$N_{\rm H}$ alone, which is $(7.3\pm 1.1)\times 10^{21}$~cm$^{-2}$, derived using the $E_{\rm B-V}$ value from MS04 and the ISM $N_{\rm H}$-$E_{\rm B-V}$
relationship given by Predehl \& Schmitt (\cite{predehl95})\footnote{To obtain a conservative error for the $N_{\rm H}-A_{\rm V}$ relationship given in 
Predehl \& Schmitt (\cite{predehl95}), we use here the
dispersion and not the error of the mean.}. 
Taking into account the optical and \xmm\ $N_{\rm H}$ 
values and error bars, we derive lower- and upper-limits of
$\approx 4.1\times 10^{21}$ and $6.7\times 10^{21}$~cm$^{-2}$ for the interstellar and total $N_{\rm H}$, respectively. This yields an
upper-limit for the intrinsic $N_{\rm H}$ of $\approx 2.6\times 10^{21}$~cm$^{-2}$.
In addition, the \xmm\ $N_{\rm H}$ values around phases 0.5 and 0.0 are compatible with being constant, unlike what may
be expected given the orbital eccentricity and inclination of the system (see Sect.~\ref{intro}), which should introduce orbital variations in the $N_{\rm H}$
intrinsic to the wind. The $3\sigma$-error of the difference of $N_{\rm H}$ between apastron and periastron is $\approx 10^{21}$~cm$^{-2}$, which we adopt as an
upper-limit to the intrinsic $N_{\rm H}$ variation between these two orbital phases. 

To quantify whether the contribution of the stellar wind to the total $N_{\rm H}$ depending on the X-ray source
location, we have modeled the wind as spherically symmetric. We have neglected
X-ray irradiation effects since the modest X-ray luminosity of the source ($\la 10^{34}$~erg~s$^{-1}$) 
is unlikely to ionize a large fraction of the wind, which could reduce the intrinsic $N_{\rm H}$ value.  
We estimate the distance at which X-rays would strongly ionize the wind to be $\la 1/10$ the orbital one
(see, e.g., Blondin \cite{blondin94}). MS04 obtained similar constraints on the ionized wind region
size. The radial wind velocity at a distance $r$ from the star has been assumed to follow
the law: $V_{\rm w}=V_{\infty}(1-R_*/r)^{\beta}$, where $V_{\infty}=2440$~km~s$^{-1}$ is the radial
velocity of the wind in the infinity,  $\beta=0.8$ is the wind profile exponent,  $R_*=9.3$~R$_{\odot}$
is the stellar radius and $r$ is the distance to the star (MS04; C05). The stellar mass-loss rate is
taken to be in the range $\dot{M}_{\rm w}\sim 3.7$--$7.5\times 10^{-7}$~M$_{\odot}$~yr$^{-1}$ (C05),
which allows us to compute the wind density: $\rho_{\rm w}=\dot{M}_{\rm w}/4\pi r^2 V_{\rm w}$. 

Integrating the wind density from the compact object location along the observer line of sight up to distances where the wind  influence is negligible, we
obtain an intrinsic $N_{\rm H}$ in the ranges $\sim 8$--$29\times 10^{21}$ and $\sim 2$--$5\times 10^{21}$~cm$^{-2}$ around phases 0.0 an 0.5, respectively.
To look for these minimum and maximum values of the wind $N_{\rm H}$, we have adopted the most extreme values for the inclination and $\dot{M}_{\rm w}$. 
The wide range of possible values comes from the factor of 2 difference in the mass-loss rate and from the two extreme inclination angles.
The
$N_{\rm H}$ difference between the two orbital locations is expected to be in the range $\sim 3$--$27\times 10^{21}$~cm$^{-2}$,  taking for this comparison
the two extreme $\dot{M}_{\rm w}$ values, due to possible orbital variability, but the same $i$  (the one rendering either the smallest or the largest
$N_{\rm H}$ difference). Now, the wide range of $N_H$ values is mainly explained by the fact that the line of sight can cross very different density
regions  at apastron and periastron passage depending on $i$.

The observed $N_{\rm H}$ seems fairly low taking into account the absorption from the ISM plus the intrinsic component computed by us. At periastron
passage, the intrinsic expected $N_{\rm H}$ value is 3 to 12 times larger than the stated upper-limit for this component. At apastron, the expected intrinsic
$N_{\rm H}$ value is similar to its upper-limit. Stronger disparities are found when comparing the calculated intrinsic $N_{\rm H}$ variation between phases
0.5 and 0.0: a factor between 3 to 27 larger than the upper-limit. It is worth noting that a similar wind model seems to work when applied, for instance, to
the orbital $N_{\rm H}$ variation in 4U~1538$-$52 (Mukherjee et~al. \cite{mukherjee07}), an X-ray binary system that presents X-ray absorption conditions
along the orbit similar to those of LS~5039. Nevertheless, there are some cases in the literature about winds in massive isolated stars that appear to be
less dense than expected (e.g. Kramer et~al. \cite{kramer03}; Cohen et~al. \cite{cohen06}). In these cases, wind clumping has been invoked to explain the low
densities of the hydrogen column density inferred from X-ray observations, although a very special wind structure would be required to prevent the observed
constancy of the $N_{\rm H}$ for such different orbital phases in LS~5039 if the emitter is well inside the system. Such special wind structure may be
produced by the slow compact object motion through the wind, although it can hardly have strong impact since the supersonic wind will efficiently convect
outwards any possible perturbation. Another reason for this low intrinsic and/or constant $N_{\rm H}$ could be that X-rays are emitted where the stellar wind
is already diluted. In particular, an emitter located along the line of sight should be beyond one orbital radius, $\approx 2\times 10^{12}$~cm, away from
the compact object to explain the constancy of the $N_{\rm H}$ or its low intrinsic value. Different orientations of the system/emitter/observer geometry
give similar constraints, thus we adopt $\sim 10^{12}$~cm as a lower-limit for the emitter-compact object distance. Finally, to reduce the inferred 
$N_{\rm H}$, the X-ray emitter may be as
large as the whole system, although in such a case the stellar wind would probably be affected at much larger extent than it seems to be (e.g. via
ionization).

\subsection{The non-thermal emitter}

The unabsorbed X-ray spectrum of LS~5039 is well explained by a pure power-law pointing to a non-thermal origin, and the data at higher energies
obtained by \inte\ (De Rosa et~al. \cite{derosa06}) show no hints of cutoff up to 100~keV, unlike it would be expected in a thermal comptonization scenario.
A location of the X-ray source at $\ga 10^{12}$~cm from the compact object would also support a non thermal nature for the high energy emission. Plausible
mechanisms that could produce the keV--MeV spectrum could be synchrotron or inverse Compton processes, although solving this question is out of the scope of
this work. We note that the lack of thermal features in the X-ray spectrum has been the main argument against accretion as the powering source in LS~5039,
and a young non-accreting pulsar, producing the high energy emission via stellar/pulsar wind collision, has been proposed as the non thermal
emitter (see, e.g., M05; Dubus \cite{dubus06}).

The emission at TeV energies in LS~5039 is unlikely produced well inside the system since photon-photon absorption does not seem to be the only effect
modulating the TeV emission (Aharonian et~al. \cite{aharonian06}). In addition, the optically thin nature of the radio emission at least down to $\sim 1$~GHz
(Mart\'i et~al. \cite{marti98}), and the relatively small radio variability of the core ($\la 50$\%), preclude a too compact radio emitter. We note that all
this fits well in a scenario in which the X-ray emitter is at some distance from the compact object. Extended radio emission has been imaged in the range
$\sim 1$--$100$~mas, showing an approximately steady orientation of $\pm 15^{\circ}$ at different epochs and spatial scales as well as strong collimation 
(Paredes et~al. \cite{paredes00,paredes02}). It seems reasonable to associate such extended radio structure with the multiwavelength emitter. Whether this
jet-like structure is powered by a different accretion mechanism, whatever its radiative properties, or it can be explained in the context of a more general
young non-accreting pulsar scenario, remains an open question. In any case, understanding the complex radio, X-ray and TeV spectra and lightcurves, and the
unclear X-rays-stellar wind link, requires further and accurate data plus modeling.

\section{Summary}

The 2005 \xmm\ observations of LS~5039 presented here were carried out around apastron and periastron passages during an epoch of enhanced stellar mass loss
rate. The fluxes, photon indices,
and $N_{\rm H}$  are $(1.18^{+0.03}_{-0.07})$ and $(0.74\pm 0.07)\times 10^{-11}$~erg~cm$^{-2}$~s$^{-1}$, 1.51$\pm 0.02$ and 1.59$\pm 0.03$,
and $(6.3\pm 0.1)$ and $(6.2\pm 0.2)\times 10^{21}$~cm$^{-2}$, at orbital phases $\approx  0.5$ and 0.0, respectively. The reanalysis of 2005
\cha\ observations yields a flux of $(2\pm 1)\times 10^{-11}$~erg~cm$^{-2}$~s$^{-1}$, a photon index of 1.1$\pm 0.2$, and $N_{\rm
H}=(6.2\pm 0.2)\times 10^{21}$~cm$^{-2}$. The \xmm\ fluxes and photon indices show variability (more marginal for the latter), whereas the
$N_{\rm H}$ is compatible with being constant. \cha\ data point to a hard spectrum, and possible higher fluxes, although the impact of pileup
adds uncertainty to these results. Collected \Halp\ EW data from 1992 to nowadays show that this quantity varies on timescales longer, and
perhaps also shorter, than one year. Comparison of the \Halp\ EW at the epochs when 2003 and 2005 \xmm\ fluxes are available (at similar
orbital phases), point to a quite constant X-ray flux versus significant \Halp\ EW variations. If confirmed, this would disfavor a wind
accretion scenario to explain the X-ray emission in the source. The constancy of the $N_{\rm H}$
points either to an emitter located at $\ga 10^{12}$~cm from the compact object or to a stellar wind densities between 3 to 27 times lower
than those expected from a spherically symmetric stellar wind model, since ionization does not seem to affect significantly the wind
absorption. An upper-limit on the intrinsic $N_{\rm H}$ is also derived, being between 3 to 12 times smaller the one predicted by the
mentioned stellar wind model at periastron passage. We propose that the multiwavelength emitter is not located inside the binary system,
probably being the jet-like structure observed at radio frequencies. 
In any case, simultaneous X-ray/optical observations are still required to clarify the stellar wind/X-ray link, and further \xmm\ or \cha\ X-ray
observations are also necessary to further investigate the flux, photon index and intrinsic $N_{\rm H}$ evolution along the orbit.

\begin{acknowledgements} 
We are grateful to M.~V. McSwain for providing us with \Halp\ EW information obtained during years 1998, 1999, 2000 and 2003. We thank Dieter Horns and Jennifer R.
West for fruitful discussions concerning the analysis of the \cha\ data presented here. We thank Patrick Guillout, Olivier Herent, Rodrigo P. Campos, M.V.F.
Copetti, Laerte Andrade, Adriana Mancini-Pires and Monica M.M. Uchida for conducting part of the optical observations at various observatories. 
V.B-R. thanks the Max-Planck-Institut f\"ur Kernphysik for its support and kind hospitality.  
V.B-R. gratefully acknowledges support from the Alexander von Humboldt
Foundation.
M.R. is being supported by a
Juan de la Cierva fellowship from the Spanish Ministerio de Educaci\'on y Ciencia (MEC). V.B-R., M.R. and J.M.P acknowledge support by DGI of MEC under grant
AYA2004-07171-C02-01, as well as partial support by the European Regional Development Fund (ERDF/FEDER).  I.N. acknowledges support by DGI of MEC under grant
AYA2005-00095. R.L.O. acknowledges financial support from the Brazilian agency FAPESP (grant 03/06861-6). This work is based on observations obtained with XMM-{\it
Newton}, an ESA science mission with instruments  and contributions directly funded by ESA Member States and NASA. 
\end{acknowledgements}

{}


\begin{thebibliography}{}

\bibitem[2005]{aharonian05}
Aharonian, F., Akhperjanian, A.~G., Aye, K.~M., et~al. 2005, Science, 309, 746

\bibitem[2006]{aharonian06}
Aharonian, F., Akhperjanian, A.~G., Bazer-Bachi, A.~R., et~al. 2006, A\&A, 460, 743

\bibitem[1992]{balu92} 
Balucinska-Church, M., \& McCammon, D.\ 1992, \apj, 400, 699 

\bibitem[1994]{blondin94}
Blondin, J.~M. 1994, ApJ, 435, 756
 
\bibitem[2005]{bosch05}
Bosch-Ramon, V., Paredes, J.~M., Rib\'o, M., et~al. 2005, \apj, 628, 388 (BR05)

\bibitem[2005]{casares05}
Casares, J., Rib\'o, M., Ribas, I., et~al. 2005, MNRAS, 364, 899 (C05)

\bibitem[2006]{cohen06} 
Cohen, D.~H., Leutenegger, M.~A., Grizzard, K.~T., et~al. 2006, MNRAS, 368, 1905

\bibitem[2001]{davis01} 
Davis, J.~E.\ 2001, ApJ, 562, 575

\bibitem[2006]{derosa06}
De Rosa, A., Ubertini, P., Bazzano, A., et~al. 2006, AdSpR, proceedings from COSPAR 2006, Ed. Vink, J. \& Ghavamian, P., 
in press [astro-ph/0611114]

\bibitem[2006]{dubus06}
Dubus, G. 2006, A\&A, 456, 801

\bibitem[2006]{goldoni06}
Goldoni, P., Rib\'o, M., Di Salvo, T., et~al. 2006, Ap\&SS, in press [astro-ph/0609708]

\bibitem[2003]{hinkle03} 
Hinkle, K.~H., Wallace, 
L., \& Livingston, W.\ 2003, Bulletin of the American Astronomical Society, 
35, 1260 

\bibitem[2006]{horns06}
Horns, D. for the HESS collaboration, 2nd Workshop On TeV Particle Astrophysics, Madison, August 2006

\bibitem[1999]{kaufer99} 
Kaufer, A., Stahl, O., 
Tubbesing, S. et~al.\ 1999, The Messenger, 95, 8 

\bibitem[2003]{kramer03}
Kramer, R.~H., Cohen, D.~H., \& Owocki, S.~P. 2003, ApJ, 592, 532

\bibitem[1990]{lemaitre90} 
Lemaitre, G., Kohler, D., Lacroix, D., Meunier, J.~P., \& Vin, A.\ 1990, \aap, 228, 546 

\bibitem[1998]{marti98}
Mart{\'{\i}}, J., Paredes, J.~M., \& Rib\'o, M. 1998, A\&A, 338, L71

\bibitem[2005]{martocchia05}
Martocchia, A., Motch, C., \& Negueruela, I. 2005, A\&A, 430, 245 (M05)

\bibitem[2002]{mcswain02}
McSwain, M.~V., \& Gies, D.~R. 2002, \apj, 568, L27

\bibitem[2004]{mcswain04}
McSwain, M.~V., Gies, D.~R., Huang, W., et~al. 2004, ApJ, 600, 927 (MS04)

\bibitem[1997]{motch97}
Motch, C., Haberl, F., Dennerl, K., Pakull, M., \& Janot-Pacheco, E. 1997, A\&A, 323, 853

\bibitem[2007]{mukherjee07}
Mukherjee, U., Raichur, H., Paul, B., Naik, S., \& Bhatt, N. 2007, [astro-ph/0702142]

\bibitem[2000]{paredes00}
Paredes, J.M., Mart\'{\i}, J., Rib\'{o}, M., \&  Massi, M. 2000, Science, 288, 2340

\bibitem[2002]{paredes02}
Paredes, J.~M., Rib\'o, M., Ros, E., Mart\'i, J., \& Massi, M. 2002, A\&A, 393, L99

\bibitem[1995]{predehl95}
Predehl, P., \& Schmitt, J.~H.~M.~M. 1995, A\&A, 293, 889

\bibitem[1996]{puls96}
Puls, J., Kudritzki, R.~P., Herrero, A., et al. 1996,  A\&A, 305, 171

\bibitem[2003]{reig03}
Reig, P., Rib\'o, M., Paredes, J.~M., \& Mart\'{\i}, J. 2003, A\&A, 405, 285 (R05)

\bibitem[1999]{ribo99} 
Rib\'o, M., Reig, P., Mart\'i, J., \& Paredes, J.~M. 1999, A\&A, 347, 518

\end{thebibliography}
\end{document}